\documentstyle[psfig]{basi}
\begin{document}
\title[SN 2002ap, the hypernova of class Ic ]{SN 2002ap, the hypernova of class Ic }

\author[S. B. Pandey et al.]%
       {S. B. Pandey$^1$\thanks{E mail: shashi@upso.ernet.in}, D. K. Sahu$^2$, G. C. Anupama$^3$, D. Bhattacharya$^4$ 
and Ram Sagar$^{1,3}$  \\
$^1$ State Observatory, Manora Peak Naini Tal -- 263 129, India\\
$^2$ Center for Research \& Education in Science \& Technology, Hoskote, Bangalore -- 562
114, India\\
$^3$ Indian Institute of Astrophysics, Bangalore -- 560 034, India\\
$^4$  Raman Research Institute, Bangalore -- 560 080, India \\}
\pubyear{2003}
\date{Received ---; accepted ---}
\maketitle
\label{firstpage}

\vspace{-0.5cm}

\begin{abstract}

The supernova SN 2002ap was discovered in the outer regions of the nearby
spiral M74 on January 29.4 UT. Early photometric and spectroscopic observations 
indicate the supernova belongs to the class of Ic hypernova.  
Late time (After JD 2452500) light curve decay slopes are similar to that of 
the hypernovae SN 1997ef and SN 1998bw. We present here the $BVRI$ photometric 
light curves and colour evolutions of SN 2002ap to investigate the late time nature 
of the light curve.

\end{abstract}

\vspace{-0.5cm}


\vspace{-0.5cm}

\section{Introduction}

\vspace{-0.5cm}

Hypernovae of class Ic are characterized by smooth and featureless spectrum
at early epochs, with no lines due to hydrogen and helium in their spectra.
According to the current interpretation, these objects have the usual lines
of SN Ic but with extreme Doppler broadening caused by unusually high
expansion velocities. The maximum brightness of SN 2002ap was $M_v = -17.2$ 
mag comparable to that of SN 1997ef but fainter than SN 1998bw 
(Patat et al. 2001, Pandey et al. 2003). Spectroscopically also, SN 2002ap 
was found to be closer to SN 1997ef than to SN 1998bw (Mazzali et al. 2002). 
The late time light curve of SN 2002ap indicates a decay rate of $\sim$ 0.016
mag day$^{-1}$, around $\sim$ 300 days after the explosion. During a similar
time interval SN 1998bw also decayed with the same rate.

\vspace{-0.5cm}

\begin{figure}
\hspace*{2.0cm}
\psfig{file=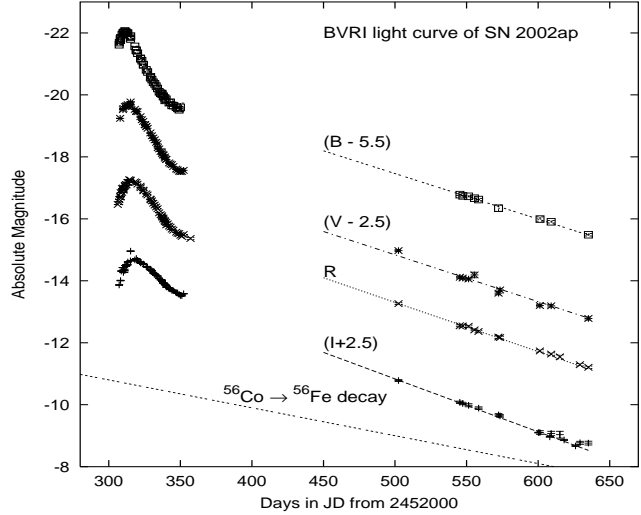,height=7cm,width=12cm,angle=-90}
\caption{\label{light} $BVRI$ Photometric Light curves of SN 2002ap Hypernova.}
\end{figure}


\section{Late Time  $BVRI$ Light Curve of SN 2002ap}

\vspace{-0.3cm}

$BVRI$ photometric observations were carried out from 104--cm NainiTal and
2--m Himalayan Chadra Telescope (HCT), Hanle using CCD cameras.
Fig. 1 shows the $BVRI$ light curve of SN 2002ap including late time observations.
The span of our present observations are from $\sim$ 2452500 JD to $\sim$
2452640 JD, similar to that in SN 1998bw (Galama et al. 1998, Patat et al. 2001).
The values of late time decay slopes are 0.0173$\pm$0.0005, 0.0158$\pm$0.0002,
0.0151$\pm$0.0005 and 0.0147$\pm$0.0002 for $I,R,V$ and $B$ bands respectively for 
the phase range of 200 -- 330 days from $B$ maxima. These values match with the 
corresponding slopes of SN 1998bw for the phase range 40 -- 330 days from $B$ maxima 
(Patat et al. 2001). If we consider the phase range  40 -- 330 days from $B$ maxima 
to determine the slope values, they are steeper than the slope values of SN 1998bw 
for the same phase range (Patat et al. 2001, Pandey et al. 2003).

\vspace{-0.5cm}

\section{ Conclusions}

\vspace{-0.5cm}

$BVRI$ photometric results are presented to investigate the light curve of SN 2002ap 
. The determined late time decay slope values for $B,V,R$ and $I$ passbands for the 
phase range 200 -- 340 days from $B$ maxima for SN 2002ap, indicate that the values 
are almost the same in all filters. This behaviour is similar to that observed in 
SN 1998bw during the same evolutionary phase (Patat et al. 2001). The (B - V), (V - R) 
and (R - I) colour evolutions show no considerable changes in the values during the 
late time, indicating a slow evolution of the late time spectrum. In the light of above, 
we conclude that SN 2002ap has not yet reached the evolutionary stage powered by 
$ ^{56}$Co $ \rightarrow ^{56}$Fe decay.

\label{lastpage}
\end{document}